\documentclass[12pt]{article}
\usepackage{amssymb}

\usepackage{a4}
\usepackage{amsmath}
\usepackage{amsfonts}
\usepackage{amsmath}
\usepackage{latexsym}
\usepackage{amsfonts}
\begin{document}

\noindent hep-th/
   \hfill  April  2008 \\

\renewcommand{\theequation}{\arabic{section}.\arabic{equation}}
\thispagestyle{empty}
\vspace*{-1,5cm}
\noindent \vskip3.3cm

\begin{center}
{\Large\bf Ultraviolet behaviour of higher spin gauge field propagators and one loop mass renormalization }

{\large Ruben Manvelyan ${}^{\dag\ddag}$, Karapet Mkrtchyan${}^{\dag\ddag}$ \\and Werner R\"uhl
${}^{\dag}$}
\medskip

${}^{\dag}${\small\it Department of Physics\\ Erwin Schr\"odinger Stra\ss e \\
Technical University of Kaiserslautern, Postfach 3049}\\
{\small\it 67653
Kaiserslautern, Germany}\\
\medskip
${}^{\ddag}${\small\it Yerevan Physics Institute\\ Alikhanian Br.
Str.
2, 0036 Yerevan, Armenia}\\
\medskip
{\small\tt manvel,ruehl@physik.uni-kl.de; karapet@yerphi.am}
\end{center}\vspace{2cm}

\bigskip
\begin{center}
{\sc Abstract}
\end{center}
The ultraviolet singular structure of the
bulk-to-bulk propagators for higher spin gauge fields in $AdS_{4}$
space is analyzed in details. Possible interactions with the Higgs scalar and
the corresponding one loop mass renormalization are studied.
This mass renormalization is finite and connected with the anomalous dimensions
of those currents in the corresponding boundary $CFT_{3}$ that cease to be conserved when the interaction is switched on. In particular it is proportional to $\ell-2$.

\newpage

\section{Introduction}
If $AdS_{4}/CFT_{3}$ correspondence \cite{Kleb} is valid beyond the tree graph approximation (classical field theory)
and supersymmetry is not presumed to hold, the anomalous dimensions of currents in $CFT_{3}$
that are conserved only in the free field limit on the one hand, and the masses of higher spin gauge fields on $AdS_{4}$ space on the other hand must be related in an accessible way. Since the anomalous dimensions of $CFT_{3}$ are derived by renormalized perturbation theory, an analogous approach,
namely one involving renormalization, ought to be possible for higher spin field theory on $AdS_{4}$ space, too. Accessibility means in particular that this renormalization of mass
is finite for all loop graphs involved.

We consider one specific loop graph for a given higher spin field, constructed from a gauge invariant vertex that involves the Higgs field. We have chosen it such that its boundary limit renormalizes the current two-point function in $CFT_{3}$. However, we expect that several different couplings and loop graphs contribute to the mass. Thus in the best case our result  besides being finite should
show a structure typical for such mass, but its numerical value could possibly disagree. In $CFT_{3}$ some currents remain conserved if the coupling is switched on. Their anomalous dimension remains zero. If these currents are characterized by a parameter that, if it takes a specific value, the current is a conserved one, then some "analyticity" presumed, the anomalous dimension should contain a factor which vanishes at this value. The simplest example is the stress-energy tensor with spin $\ell=2$. So we expect the anomalous dimension and consequently also the higher spin field mass to contain the factor $\ell-2$.

The $AdS$ field theory we study lives in $AdS_{4}$ and is of Fronsdal type \cite{Frons}, namely the higher spin fields have vanishing double trace. We can restrict ourselves on models with only even spin $\ell$
and with one scalar (Higgs) field. Its boundary is supposed to be the conformal field theory consisting of all $O(N)$ invariant local fields in the $O(N)$ conformal sigma model $CFT_{3}$
\cite{Kleb},\cite{LMR3}.
The whole field algebra is assumed to be closed under operator product expansions. A corresponding extension of the higher spin field theory is also assumed. This kind of $CFT_{3}$
has the advantage of possessing a renormalizable and simple perturbative expansion ($1/N$ expansion) and that the anomalous dimensions of an infinite number of currents that are conserved in the free field limit (and then have dimensions $d+\ell-2, d=3$) have been computed to first order \cite{Ruehl}
\begin{equation}
\eta_{\ell} = \frac{1}{N}\frac{16(\ell-2)}{3\pi^{2}(2\ell-1)} ,
\label{0.0}
\end{equation}
implying the mass
\begin{equation}
\delta m_{\ell}^{2} = \frac{1}{N}\frac{16(\ell-2)}{3\pi^{2}} ,
\label{0.1}
\end{equation}
for the higher spin field of spin $\ell$.

In a sense this is the simplest pair of candidates for $AdS_{4}/CFT_{3}$ correspondence we can
think of. We do not wish to entangle it with complications caused by extra fields which leave their traces at least in off shell amplitudes and thus may obscure simple structures.

In Section 2 we provide all tools and concepts necessary for the investigations in the subsequent parts of this work. These methods have been selected or developed
in a series of former works of the authors \cite{MR1}, \cite{MR2}, \cite{MR3}, \cite{MR4}, \cite{MR5}, \cite{MR6}. There exists a basic gauge in which higher spin quantum field theory can be formulated: de Donder's gauge. In Section 3 we present an extensive treatment of the higher spin propagators in this gauge and in both the ultraviolet (UV) and the infrared (IR) domains, which have a common intersection of convergence and analyticity.
Nevertheless we lack a method for explicit analytic continuation from one domain into the other,
since the functions arising are apparently not of the generalized hypergeometric class. A possible solution to this problem is to go over to another gauge (Feynman's gauge, preferentially) do the analytic continuation there, and then invert the gauge transformation.

The existence of Feynman's gauge for any higher spin field is proved in Section 4. In this gauge the propagators are Gaussian hypergeometric and the analytic continuations are standard.
Feynman's gauge is obviously the choice for any perturbative treatment of AdS-QFT.
However, to preserve gauge invariance propagators must couple to covariantly conserved
currents. In Section 5 such currents formed from one higher spin field of
spin $\ell$ and the Higgs scalar and aimed to be coupled to another higher spin field
of spin $\ell\pm 2$ are constructed. These currents appear as cosets with respect to the
dual of the de Donder gauge operator which is a novel constructive concept.

The loop function belonging to two such vertices has two higher spin fields as external legs
that are "amputated" in the language of QFT: they are operated on with the Fronsdal operator
which makes them gauge invariant, and are off-shell. Internally we have the product (in coordinate space) of the scalar and the trace of the spin $\ell+2$ field in Feynman
gauge. There is a logarithmic UV singularity which can be regularized and renormalized
by subtraction of an infinite gauge invariant counter term and a finite mass renormalization
expressed by (\ref{5.24}) which is our main result.

\setcounter{equation}{0}
\section{General setup for higher spin propagators}

Here we would like to gather all technical details developed in our different articles about several problems of $AdS_{d+1}$ higher spin theories\footnote{We will always try to keep general $d$ in all possible formulas admitting of course that for our $AdS_{4}$ theory it should be set to $3$ at the end }.
We work in Euclidian $AdS_{d+1}$ with the following metric,
curvature and covariant derivatives:
\begin{eqnarray}
&&ds^{2}=g_{\mu \nu }(z)dz^{\mu }dz^{\nu
}=\frac{L^{2}}{(z^{0})^{2}}\delta _{\mu \nu }dz^{\mu }dz^{\nu
},\quad \sqrt{g}=\frac{L^{d+1}}{(z^{0})^{d+1}}\;,
\notag  \label{ads} \\
&&\left[ \nabla _{\mu },\,\nabla _{\nu }\right] V_{\lambda }^{\rho }=R_{\mu
\nu \lambda }^{\quad \,\,\sigma }V_{\sigma }^{\rho }-R_{\mu \nu \sigma
}^{\quad \,\,\rho }V_{\lambda }^{\sigma }\;,  \notag \\
&&R_{\mu \nu \lambda }^{\quad \,\,\rho
}=-\frac{1}{(z^{0})^{2}}\left( \delta _{\mu \lambda }\delta _{\nu
}^{\rho }-\delta _{\nu \lambda }\delta _{\mu }^{\rho }\right)
=-\frac{1}{L^{2}}\left( g_{\mu \lambda }(z)\delta _{\nu
}^{\rho }-g_{\nu \lambda }(z)\delta _{\mu }^{\rho }\right) \;,  \notag \\
&&R_{\mu \nu }=-\frac{d}{(z^{0})^{2}}\delta _{\mu \nu }=-\frac{d}{L^{2}}%
g_{\mu \nu }(z)\quad ,\quad R=-\frac{d(d+1)}{L^{2}}\;.  \notag
\end{eqnarray}%
For simplicity we will from now on put $L=1$ during all calculations keeping in mind that we can always restore the $AdS$ radius from dimensional consideration.
The next important technical trick is the contraction of all the rank $\ell$ symmetric tensors $h^{(\ell)}_{\mu_{1}\dots \mu_{\ell}}(z)$ with the $\ell $-fold tensor product of a
vector $a^{\mu }$
\begin{equation}\label{1.1}
 h^{(\ell)}(z;a)= h^{(\ell)}_{\mu_{1}\mu_{2}\dots \mu_{\ell}}(z)a^{\mu_{1}}a^{\mu_{2} }\dots a^{\mu_{\ell}} .
\end{equation}
In this notation the trace and divergence of the symmetric tensors are second order differential operators in $(z;a)$ ``hyperspace''
\begin{eqnarray}
  \ell(\ell-1)Trh^{(\ell)}(z;a)&=& \Box_{a}h^{(\ell)}(z;a)= g^{\mu \nu
}\frac{\partial ^{2}}{\partial a^{\mu }\partial a^{\nu }}h^{(\ell)}(z;a) ,\label{1.2}\\
  \ell \,Divh^{(\ell)}(z;a)&=& \nabla ^{\mu
}\frac{\partial }{\partial a^{\mu }}h^{(\ell )}(z;a)=(\nabla \partial_{a})h^{(\ell )}(z;a) .
\end{eqnarray}

Then we can write the starting point of the investigation into higher spin gauge field propagators, namely Fronsdal's equation of motion (we introduce here the ``geometric AdS mass'' $\mu^{2}_{\ell}$)
\cite{Frons} for the double traceless spin $\ell$ field, $\Box _{a}\Box _{a}h^{(\ell )}=0$:
\begin{eqnarray}
\mathcal{F}(h^{(\ell )}(z;a))&=&[\Box-\mu^{2}_{\ell}] h^{(\ell )}(z;a)-a^{2}\Box
_{a}h^{(\ell)}(z;a)\quad   \nonumber  \\
&-&(a\nabla )\Big[\nabla ^{\mu
}\frac{\partial }{\partial a^{\mu }}h^{(\ell )}-\frac{1}{2}(a\nabla
)\Box _{a}h^{(\ell )}(z;a)\Big]=0 ,\label{1.4}\\
\mu^{2}_{\ell}&=&\left( \ell ^{2}+\ell (d-5)-2(d-2)\right) ,  \label{1.5} \\
\Box&=&\nabla^{\mu}\nabla_{\mu},\quad
(a\nabla)=a^{\mu}\nabla_{\mu},\quad
a^{2}=g_{\mu\nu}(z)a^{\mu}a^{\nu} .\label{1.6}
\end{eqnarray}%
The most important property of this equation is higher spin gauge invariance with the
traceless parameter $\epsilon ^{(\ell -1)}(z;a),$
\begin{eqnarray}
\delta h^{(\ell )}(z;a)&=&(a\nabla )\epsilon ^{(\ell -1)}(z;a),\label{1.7}\\
\Box_{a}\epsilon ^{(\ell -1)}(z;a)&=&0,\label{1.8}\\
\delta \mathcal{F}(h^{(\ell)}(z;a))&=&0.  \label{1.9}
\end{eqnarray}%
supplemented with the "Bianchi" identity
\begin{equation}\label{1.10}
    (\nabla\partial_{a})\mathcal{F}-\frac{1}{2}(a\nabla)\Box_{a}\mathcal{F}=0 .
\end{equation}
Comparing (\ref{1.10}) with the second line of (\ref{1.4}) and keeping in mind \emph{tracelessness} of the gauge parameter (\ref{1.7}), we arrive at the idea that the most natural gauge fixing condition for Fronsdal's equation is the so called \emph{traceless} de Donder gauge
\begin{eqnarray}
&&\mathcal{D}^{(\ell -1)}(h^{(\ell )})=\nabla ^{\mu }\frac{\partial }{%
\partial a^{\mu }}h^{(\ell )}-\frac{1}{2}(a\nabla )\Box _{a}h^{(\ell )}=0,
\label{1.11}
\end{eqnarray}
In this gauge Fronsdal's equation simplifies to
\begin{eqnarray}
&&\mathcal{F}^{dD}(h^{(\ell )})=[\Box-\mu^{2}_{\ell}] h^{(\ell )}-a^{2}\Box _{a}h^{(\ell)}=0.\quad \quad \label{1.12}
\end{eqnarray}
Note that any deviation in (\ref{1.12}) of $\mu^{2}_{\ell}$ from (\ref{1.5}) leads to a massive higher spin field.

A further useful notation is an abbreviated contraction of tensor indices
\begin{eqnarray}
  *_{a}&=&\frac{1}{(\ell!)^{2}} \prod^{\ell}_{i=1}\overleftarrow{\partial}^{\mu_{i}}_{a}\overrightarrow{\partial}_{\mu_{i}}^{a} .
   \label{1.13}
\end{eqnarray}
Then we see that operators $a^{2}$ and $\Box_{a}$ are dual (or adjoint)  with respect to the "star" product (\ref{1.13}) in the same fashion as $(a\nabla)$ and $(\nabla\partial_{a})$ are
dual (or adjoint) with respect to the full scalar product in the space $(z;a)$
\begin{eqnarray}
    a^{2}f(a)*_{a}g(a)&=&f(a)*_{a}\Box_{a}g(a) , \label{1.14}\\
    \int\sqrt{g}d^{4}z (a\nabla)f(z;a)*_{a}g(z;a) &=& -\int\sqrt{g}d^{4}z f(z;a)*_{a}(\nabla\partial_{a})g(z;a) .\label{1.15}
\end{eqnarray}
Thus we can write now Fronsdal's  gauge invariant action in the concise form
\begin{eqnarray}
  &&\mathcal{S^{\ell}} = \frac{1}{2} \int\sqrt{g}d^{4}z\left\{h^{(\ell )}(z;a)*_{a}\mathcal{F}(h^{(\ell )}(z;a))-\frac{1}{4}\Box_{a}h^{(\ell )}(z;a)*_{a}\Box_{a}\mathcal{F}(h^{(\ell )}(z;a))\right\} .\quad\quad\quad\label{1.16}
\end{eqnarray}
This action is gauge invariant due to the "Bianchi" identity (\ref{1.10}).

Next we can write our double traceless field $h^{(\ell)}(z;a)$ as a set of the two traceless spin $\ell$ and $\ell-2$
fields $\psi^{(\ell)}(z;a)$ and $\theta^{(\ell-2)}(z;a)$
\begin{eqnarray}
&& h^{(\ell)}(z;a)= \psi^{(\ell)}+\frac{a^{2}}{2\alpha_{0}}
\theta^{(\ell-2)}(z;a)\quad,  \label{1.17} \\
&&\Box_{a}h^{(\ell)}=\theta^{(\ell-2)}\quad,\quad
\Box_{a}\psi^{(\ell)}=\Box_{a}\theta^{(\ell-2)}=0 ,\label{1.18}\\
&&\alpha_{0}=d+2\ell-3 .\label{1.19}
\end{eqnarray}
Applying the de Donder gauge condition we see that the fields $\psi^{(\ell)}(z;a)$ and $\theta^{(\ell-2)}(z;a)$ completely separate in the action (\ref{1.16})
\begin{eqnarray}
  \mathcal{S^{\ell}} &=& \frac{1}{2} \int\sqrt{g}d^{4}z\Big\{\psi^{(\ell )}(z;a)*[\Box-\mu^{2}_{\ell}]\psi^{(\ell )}(z;a) \nonumber\\
  &-&\frac{\alpha_{0}-2}{4\alpha_{0}}\theta^{(\ell-2)}(z;a)*[\Box-\mu^{2}_{\theta^{(\ell-2)}}]\theta^{(\ell-2)}(z;a)\Big\} ,\label{1.20}\\
  \mu^{2}_{\theta^{(\ell-2)}}&=& \mu^{2}_{\ell}+2\alpha_{0}=\ell(\ell+d-1)-2 ,\label{1.21}
\end{eqnarray}
with the following diagonal field equations and de Donder gauge condition connecting  $\psi^{(\ell)}$ and $\theta^{(\ell-2)}$
\begin{eqnarray}
&& \nabla^{\mu}\frac{\partial}{\partial a^{\mu}}\psi^{(\ell)}=
\frac{\alpha_{0}-2}{2\alpha_{0}}(a\nabla)\theta^{(\ell-2)}-\frac{a^{2}}{%
2\alpha_{0}} \nabla^{\mu}\frac{\partial}{\partial
a^{\mu}}\theta^{(\ell-2)},\label{1.22} \\
&&\left(\Box+\ell\right)\psi^{(\ell)}=\Delta_{\ell}(\Delta_{\ell}-d)
\psi^{(\ell)} ,\qquad  \label{1.23} \\
&&\left(\Box+\ell-2\right)\theta^{(\ell-2)}=\Delta_{\theta}(\Delta_{%
\theta}-d)\theta^{(\ell-2)} ,
\label{1.24} \\
&&\Delta_{\ell}=d+\ell-2\quad,\quad \Delta_{\theta}=d/2+1/2\sqrt{(\alpha_{0}-1)(\alpha_{0}+7)} .\label{1.25}
\end{eqnarray}
So we realize that only in the de Donder gauge we have a diagonal
equation of motion for the physical $\psi^{(\ell)}$ components but
this component is not transversal due to the presence of
$\theta^{(\ell-2)}$ . In our previous paper \cite{MR1} we have shown that this gauge is
most convenient for the quantization and construction of
bulk-to-bulk propagators and for the investigation of
$AdS_{4}/CFT_{3}$ correspondence in the case of the critical
conformal $O(N)$ boundary sigma model. We also mentioned that in the boundary limit only the traceless mode $\psi^{(\ell)}$ survives but the nonphysical trace mode $\theta^{(\ell-2)}$ can create a Goldstone mode and enters the bulk tree dynamics and the loops.

The negative sign of the $\theta$ part in the action (\ref{1.16}) suggests to quantize this
higher spin field theory with a formalism of Gupta-Bleuler type, so that a state with $n$ quanta of $\theta$ has norm squared of signature $(-1)^{n}$ yielding a pseudo Hilbert space. Applying de Donder's constraint (\ref{1.22}) on field operators, the ``physical'' Hilbert space is the kernel of the annihilation operator part of this constraint inside the pseudo Hilbert space. Finally zero norm states are projected out. In the context of this work it is only
relevant that the two-point function of $\theta$ satisfies
\begin{equation}
\big<\theta^{(\ell-2)}(z_1;a),\theta^{(\ell-2)}(z_2;c)\big> \quad \leq \quad 0
\label{1.26}
\end{equation}
as a distribution.

\setcounter{equation}{0}
\section{Propagators in de Donder's gauge}
On AdS space which is a constant curvature space the geodesic distance $\eta$ enters all invariant expressions of the relative distance of two points. The standard variable $\zeta = \cosh \eta$ can be expressed by Poincar\'e coordinates
as
\begin{equation}
        \zeta(z_{1},z_{2})=\frac{(z^{0}_{1})^{2}+(z^{0}_{2})^{2}+(\vec{z}_{1}-\vec{z}_{2})^{2}}
        {2z^{0}_{1}z^{0}_{2}}=1+\frac{(z_{1}-z_{2})^{\mu}(z_{1}-z_{2})^{\nu}
        \delta_{\mu\nu}}{2z^{0}_{1}z^{0}_{2}} .
 \end{equation}
The propagators are bitensorial quantities which are presented in the algebraic basis
of homogeneous functions of $I_1, I_2, I_3, I_4$
\begin{eqnarray}
       && I_{1}(a,c):=(a\partial_{1})(c\partial_{2})\zeta(z_{1},z_{2}) , \\
       && I_{2}(a,c):=(a\partial_{1})\zeta(z_{1},z_{2})(c\partial_{2})\zeta(z_{1},z_{2}),\\
       && I_{3}(a,c):=a^{2}_{1}I^{2}_{2c}+c^{2}_{2}I^{2}_{1a} , \\
       && I_{4}:=a^{2}_{1}c^{2}_{2} ,\\
       && I_{1a}:=(a\partial_{1})\zeta(z_{1},z_{2})\quad ,
       \quad I_{2c}:=(c\partial_{2})\zeta(z_{1},z_{2}) , \\
       &&(a\partial_{1})=a^{\mu}\frac{\partial}{\partial
       z_{1}^{\mu}} ,\quad (c\partial_{2})=c^{\mu}\frac{\partial}{\partial
       z_{2}^{\mu}} ,\\&& a^{2}_{1}=g_{\mu\nu}(z_{1})a^{\mu}a^{\nu} ,
       \quad c^{2}_{2}=g_{\mu\nu}(z_{2})c^{\mu} c^{\nu} .
\end{eqnarray}
of degree $\ell$, the spin of the field. All important formulas for this "advanced technology" of working with higher spin field theory in $AdS$ space one can find in Appendix A.
We are interested only in that part of the propagator expansion which neglects traces. So it is a map from a space of $\ell+1$ functions $\left\{F_{k}(\zeta)\right\}_{k=0}^{\ell}$ to a space of bitensors  parameterized by $I_{1}$ and $I_{2}$ only, namely
\begin{eqnarray}
 \Psi^{(\ell)}[F_{k}]& =& \sum_{k=0}^{\ell} F_{k}(\zeta) I_{1}^{\ell-k} I_{2}^{k} ,\label{2.0}\\
 \left(\Box+\ell\right)\Psi^{(\ell)}[F_{k}]&=&\Delta_{\ell}(\Delta_{\ell}-d)\Psi^{(\ell)}[F_{k}] +O(a^{2}_{1};c^{2}_{2}) .
\end{eqnarray}

In the variable $\zeta$ the
analytic properties of QFT n-point functions are conveniently described.
In particular the two-point functions or propagators are analytic in the $\zeta$ plane with singularities at $\zeta = \pm 1$ and at $\zeta = \infty$, which in most cases are
logarithmic branch points. Analyticity is therefore meant in general on infinite covering planes. All $AdS$ field theories are symmetric under the exchange $\zeta$ against $-\zeta$.

Another variable used often is $u = \zeta -1$, the ``chordal distance'', more precisely one half the square of the chordal distance. The series expansions for two-point functions in $u$ converge in a radius $2$,
whereas the series expansions in powers of $\zeta^{-1}$ converge for $\mid\zeta\mid > 1$.
These analytic properties remind us of Legendre functions. Indeed if propagator functions can be identified as Gaussian hypergeometric functions, these are Legendre functions and the "quadratic transformations" can be applied.
Using formulas from Appendix A we can show that
in de Donder's gauge the propagator satisfy the following set of differential equations
for the functions $F_{k}(\zeta)$ or correspondingly $\Phi_{k}(u)$ of (\ref{2.0}) following from equation (\ref{1.23})
\begin{eqnarray}
(\zeta^{2} -1)F_{k}'' +(d+1+4k)\zeta F_{k}' +X_{k}F_{k} +2\zeta(k+1)^{2}F_{k+1}
+2(\ell-k+1)F_{k-1}' = 0 ,\quad\label{2.01}\\
X_{k} = k(d+2\ell-k) +2l - (\ell-2)(\ell+d-2) .\qquad\qquad\qquad
\label{2.03}
\end{eqnarray}
The "dimension" of the higher spin field $\Delta_{\ell} = \ell+d-2$
has been inserted. Moreover we use $F_{-1} = F_{\ell+1} = 0$. The dimension of the
AdS space is $d+1$, we interpolate analytically in $d$ if this is technically required.
Our issue is to solve these equations by expansion in powers of $\zeta^{-1}$ or $u$.
This leads to matrix recursion equations which necessitate some linear algebra operations.

As an ansatz for the series expansion of $F_{k}(\zeta)$ at $\zeta = \infty$ we use
\begin{equation}
F_{k}(\zeta) = \zeta^{-\alpha -k}\sum_{n=0}^{\infty} c_{kn} \zeta^{-2n} .
\label{2.1}
\end{equation}
Denote $\xi = \alpha +2n $. Then a two term recursion of the form
\begin{eqnarray}
D_{n} \left(\begin{array}{ccc}  c_{0n}\\c_{1n}\\ \vdots \\c_{\ell,n}
\end{array}\right) = C_{n-1} \left( \begin{array}{ccc} c_{0,n-1}\\ c_{1,n-1}\\
 \vdots \\ c_{\ell,n-1} \end{array} \right) ,
\label{2.2}
\end{eqnarray}
results with the two matrices
\begin{equation}
C_{n-1} = diag\{(\xi -1)(\xi -2),\xi (\xi-1), \ldots (\xi+\ell-1)(\xi +\ell-2)\} ,
\label{2.3}
\end{equation}
and the entries of the matrix $D_{n}$
\begin{eqnarray}
(D_{n})_{k,k-1} &=& -2(\ell-k+1)(\xi+k-1) ,\label{2.4}\\
(D_{n})_{k,k}  &=& \xi^{2} -\xi (d+2k) -4k^{2}+2\ell(k+1) -(l-2)(\ell+d-2) ,\label{2.5} \\
(D_{n})_{k,k+1} &=& 2(k+1)^{2} .\label{2.6}
\end{eqnarray}
The determinant of $D_{0}$ is a polynomial of degree $2(\ell+1)$ of the variable $\alpha$
with roots which we identify with the "roots" of the differential equation system.
For arbitrary $\ell$ we have
\begin{eqnarray}
det D_{0} = [(\alpha +\ell-2)(\alpha +2-\ell-d)][(\alpha +\ell-2)(\alpha -\ell-d)] \nonumber\\ \times \prod_{n=0}^{\ell-2}[\alpha^{2} -(d+4+2n)\alpha -((\ell-2)d + (\ell+n)^{2}-(n+2)(3n+4))] .
\label{2.7}
\end{eqnarray}
Each square bracket represents one eigenvalue of $D_{0}$ and contributes two roots. The quadratic factors lead in almost all cases to two irrational roots that are neither degenerate among themselves nor with the other roots, but there are exceptions which have two integer roots e.g. for $d=3: (\ell,n)\in \{(4,1), (6,4), (9,2), (9,5), (11,8), (15,8)... \}$. Two roots are said to be degenerate, if their difference is an integer. For the case of expansions in powers of $\zeta^{2}$ as in (\ref{2.1}), this integer must be even. In such case the solution with the bigger root enters the other one with a $log\zeta$ factor.

The following roots are of particular (physical) importance
\begin{eqnarray}
\alpha_{p} &=& \ell+d-2 ,\label{2.8}\\
\alpha_{c} &=& \ell+d .
\label{2.9}
\end{eqnarray}
We call the first root $\alpha_{p}$ "principal"
because it has the value of the dimension $\Delta$ of the field which enters the field
equation in the form $\Delta(\Delta-d)$. The second root is a "companion" of it, since they appear for all $\ell$ as such pair (see (\ref{2.7})). It is degenerate with the principal root and the solution of it
enters the principal solution with a $log \zeta$ factor on the next to leading power in the expansion. The bigger ones of the two roots in the exceptional cases quoted above are also bigger than the principal root $\ell +1$ (for the same $\ell$) but their distance to it are odd numbers except for the case $(\ell,n) = (15,8)$, where the distance to $\ell+1$ is sixteen
and the $log\zeta$ term appears at a very high power.

For the principal root the equation for the eigenvector of $D_{0}$
\begin {eqnarray}
D_{0}(\alpha_{p}) \left (\begin{array}{ccc}  c_{00}^{(\alpha_{p})}\\ c_{10}^{(\alpha_{p})}\\ \vdots \\ c_{l0}^{(\alpha_{p})} \end{array} \right) = 0 ,
\label{2.10}
\end{eqnarray}
can be solved for each $\ell$. We find
\begin{equation}
c_{k,0}^{(\alpha_{p})} = (-1)^{k} \binom{\ell}{k} ,
\label{2.11}
\end{equation}
which is easy to prove by using the general expression for the rows of the matrix $D_{n}$
as given in (\ref{2.4}) - (\ref{2.6}). The consequence of this result is that the leading term
of $\Psi^{(\ell)}[F_{k}(\alpha_{p})]$ at $\zeta = \infty$ is the well known expression $\zeta^{-\Delta}(I_{1} - \zeta^{-1}I_{2})^{\ell}$. Already at next order in $\zeta^{-2}$ log-terms appear.

For the companion root $\alpha_{c}$ the eigenvector for $D_{0}$ can be derived by
a little bit more algebra for any $\ell$
\begin{equation}
c_{k,0}^{(\alpha_{c})} = (-1)^{k}\left( \binom{\ell}{k} +(d+2\ell-2) \binom{\ell-1}{k-1} \right) .
\label{2.13}
\end{equation}
The actual construction of a solution for the pair of roots starts with the bigger one, $\alpha_{c}$. Its solution takes the form
\begin{equation}
F_{k}(\zeta; \alpha_{c}) = \zeta^{-\Delta -2} \sum_{n=0}^{\infty} \zeta^{-2n}\sum_{s=0}^{\ell}
\Pi_{n}(\alpha_{c})_{k,s}c_{s,0}^{(\alpha_{c})} ,
\label{2.14}
\end{equation}
where we used
\begin{eqnarray}
H_{n}(\alpha_{c}) &=& D_{n}(\alpha_{c})^{-1} C_{n-1}(\alpha_{c})\nonumber\\
                  &=& H_{1}(\alpha_{c} +2(n-1)) ,\label{2.15}\\
\Pi_{n}(\alpha_{c}) &=& \Pi_{r=0}^{n-1,\leftarrow}H_{1}(\alpha_{c} +2r) .
\label{2.16}
\end{eqnarray}
and the left arrow denotes ordering of the product with increasing $r$ from right to left.
In this context we note that if a nonsingular matrix $S(\alpha)$ would exist, so that $H_{1}$ could be diagonalized by
\begin{equation}
H_{1}(\alpha) = S^{-1}(\alpha +2) \Delta(\alpha) S(\alpha) ,
\label{2.17}
\end{equation}
then $F_{k}(\zeta;\alpha)$ would be a generalized hypergeometric function.

Having constructed the solution for the companion root we turn to the principal root.
We recognize that $D_{n}(\alpha_{p})$ can be spectrally decomposed in the following fashion
\begin{eqnarray}
D_{n}\chi_{i} &=& \lambda_{i}\chi_{i} ,\label{2.18}\\
D_{n}^{T}\psi_{i} &=& \lambda_{i}\psi_{i} ,\label{2.19}\\
D_{n} &=& \sum_{i=0}^{l} \lambda_{i} \chi_{i} \otimes \psi_{i}^{T} ,\label{2.20}\\
 \psi_{i}^{T} \chi_{j} &=&\delta_{ij} \label{2.21}
\end{eqnarray}
Denote further
\begin{equation}
\rho ^{T} = \psi^{T} C_{n-1} .
\label{2.22}
\end{equation}
All these quantities can be determined as functions of $\xi$, and it is easily verified that
(\ref{2.17}) is not fulfilled.

One of the eigenvalues of $D_{1}(\alpha_{p})$ vanishes, we denote it $\lambda_{0}$,
so that $D_{1}(\alpha_{p})$ cannot be inverted. We perform a deformation
of our differential equation system replacing $\alpha_{p}$ only in $\lambda_{0}$
and in the prefactor $\zeta^{-\alpha_{p}}$ by $\alpha_{p} + \epsilon$. All other eigenvalues and the eigenvectors remain unchanged. Then we continue the whole procedure known from the
companion root, all $H_{n}$ will remain singularity free. At the end we subtract a certain multiple $\gamma$ of
$(\epsilon^{-1}+\mu)\Psi^{(\ell)}[F_{k}(\alpha_{c})]$ so that the limit $\epsilon \rightarrow 0$
can be performed and the log-terms appearing are $-\gamma log\zeta \Psi^{(l)}[F_{k}(\alpha_{c})]$. The additional parameter $\mu$ is in principle arbitrary
showing that the principal solution containing a log factor is a coset with respect to adding the companion solution. This parameter can, however, be normalized in a standard fashion by requiring that the $(l+1)$-tupel of coefficients $c_{k,n}^{(\alpha_{p})}$ where at level $n$ the log term appears first, is orthogonal to the eigenvector $\psi_{0}$ belonging to the deformed eigenvalue.
We close this discussion with the remark that on the boundary of AdS space i.e. $\zeta = \infty$ any linear combination
\begin{equation}
\Psi^{(\ell)}[F_{k}(\alpha_{p})] + A \Psi^{(\ell)}[F_{k}(\alpha_{c})]
\label{2.23}
\end{equation}
is indistinguishable from the pure principal solution. Thus the boundary constraint
fixes only the whole coset and not any representative of it.

In order to render the expansions of $F_{k}$  around $\zeta =1(u=0)$ a visually different expression, we shall denote them $\Phi_{k}$. The expansions are
\begin{equation}
\Phi_{k}(u) = u^{\alpha}\sum_{n=0}^{\infty}a_{k,n}u^{n} .
\label{3.1}
\end{equation}
Again we obtain matrix recursion relations
\begin{eqnarray}
A_{n}\left( \begin{array} {ccc}a_{0,n}\\ a_{1,n}\\ \vdots\\ a_{\ell,n}\end{array} \right) +
B_{n-1} \left( \begin{array}{ccc} a_{0,n-1}\\ a_{1,n-1}\\
\vdots\\ a_{\ell,n-1}\end{array} \right) + E \left( \begin{array}{ccc}a_{0,n-2}\\a_{1,n-2}\\\vdots\\ a_{\ell,n-2}\end{array} \right)
= 0 .
\label{3.2}
\end{eqnarray}
We define
\begin{equation}
\xi = \alpha + n ,
\label{3.3}
\end{equation}
and obtain the matrices
\begin{eqnarray}
(A_{n})_{k,k} &=& \xi(2\xi +d +4k -1) ,\label{3.4}\\
(A_{n})_{k,k-1} &=& 2\xi(\ell-k+1) ,\label{3.5}\\
(B_{n-1})_{k,k} &=& (\xi-1)(\xi + d +4k -1) + X_{k} ,label{3.6}\\
(B_{n-1})_{k,k+1} &=& 2(k+1)^{2}
                  = (E)_{k,k+1} \label{3.7} .
\end{eqnarray}
Here we used the shorthand (see (\ref{2.03}))
\begin{equation}
X_{k}(\lambda) = k(2\lambda +2\ell -k +1) +2\ell -(\ell-2)(2\lambda +\ell -1) ,
\label{3.8}
\end{equation}
and $ d=2\lambda+1 $ has been introduced. Therefore $A_{n}$ is of lower triangular shape with eigenvalues $\xi(2\xi +d +4k-1)$. The root
system is therefore
\begin{itemize}
\item  $\ell+1$ times the root zero;
\item  the $\ell+1$ roots $\alpha_{m} =-\lambda -2m, \qquad 0 \leq m \leq \ell$.
\end{itemize}

Both sets are degenerate among themselves, and if $d$ is odd, the second set is degenerate with respect to the first one. The first set produces regular solutions, the second set
produces poles if $d$ is odd, which it is in the case of present interest. Nevertheless we will regard $d$ as a free real parameter in order to handle the degeneracy with the regular cases. The solution for $\alpha_{0}$ in combination with any regular solution has the appropriate singular behaviour at $u=0$ needed for a propagator, namely applying Fronsdal's differential operator the correct delta function is created.

Any solution is obtained by requiring
\begin{eqnarray}
A_{0} \left( \begin{array}{ccc} a_{0,0} \\a_{1,0}\\ \vdots\\ a_{\ell,0}\end{array} \right) = 0 .
\label{3.9}
\end{eqnarray}
This requirement is solved for the regular solutions $\Phi_{k}^{(r)}(u)$ (for which $A_{0} = 0$ and the solution is trivial) by
\begin{equation}
a_{k,0}^{(r)} = \delta_{k,r} .
\label{3.10}
\end{equation}
For any such solution $r$ we obtain next
\begin{eqnarray}
a_{k,1}^{(r)} &=& -(A_{1}^{-1}B_{0})_{k,r} \nonumber \\
              &=& -(A_{1}^{-1})_{k,r}(B_{0})_{rr} -(A_{1}^{-1})_{k,r-1}(B_{0})_{r-1,r} ,
\label{3.11}
\end{eqnarray}
where we insert
\begin{eqnarray}
(A_{1})_{r,r} &=& d+4r+1 ,\label{3.12}\\
(A_{1})_{r,r-1} &=& 2(\ell-r+1) ,
\label{3.13}\\
(B_{0})_{r,r} &=& X_{r} ,
\label{3.14}\\
(B_{0})_{r-1,r} &=& 2r^{2} ,
\label{3.15}
\end{eqnarray}
and obtain
\begin{eqnarray}
(A_{1}^{-1})_{k,r} &=& (-2)^{k-r}\prod_{s=r+1}^{k} (\ell-s+1)\quad [\prod_{s=r}^{k}(d+4s+1)]^{-1}
                        \quad    (\textnormal{for}\quad k > r) ,\label{3.16}\\
(A_{1}^{-1})_{r,r} &=& (d+4r+1)^{-1}\label{3.17} ,\\
(A_{1}^{-1})_{k,r} &=& 0 \quad\textnormal{for} \quad k<r .
\label{3.18}
\end{eqnarray}
There is no sign of any singularity caused by the degeneracy. Finally we get
\begin{equation}
a_{k,1}^{(r)} = -X_{r}(A_{1}^{-1})_{k,r} -2r^{2}(A_{1}^{-1})_{k,r-1} ,
\label{3.19}
\end{equation}
which vanishes for $r>k+1$.

We turn now to the nonanalytic solutions $\Phi_{k}(u,\alpha_{m})$ with roots $\alpha_{m} = -\lambda -2m$ and concentrate on the case $m=0$ because this is the perturbative Green function for the Fronsdal differential
operator. At the beginning we assume $\lambda \notin \textbf Z$ in order to avoid the degeneracy with the regular solutions. In this case we have
\begin{eqnarray}
(A_{0})_{k,k} &=&-4\lambda k ,\label{3.20}\\
(A_{0})_{k,k-1} &=&-2\lambda (\ell-k+1) ,
\label{3.21}
\end{eqnarray}
and the equation
\begin{equation}
\sum_{r}(A_{0})_{k,r} c_{r,0}^{(\alpha_{0})} = 0
\label{3.22}
\end{equation}
is solved by
\begin{equation}
c_{k,0}^{(\alpha_{0})} = \left(-\frac{1}{2}\right)^{k} {\ell\choose k} .
\label{3.23}
\end{equation}
Next we treat the $A_{1}$ matrix
\begin{eqnarray}
(A_{1})_{k,k} &=& 2(1-\lambda)N_{k},\quad N_{k} = 2k+1 ,\label{3.24}\\
(A_{1})_{k,k-1} &=& 2(1-\lambda)(\ell-k+1) ,\label{3.25}\\
(A_{1}^{-1})_{k,r} &=& [2(1-\lambda)]^{-1} \beta_{k,r}, \textnormal{for}\quad k\geq r
\quad \textnormal{and zero else} ,\label{3.26}\\
\beta_{k,r} &=&(-\ell)_{k-r}[\prod_{s = r}^{k} N_{s}]^{-1} .
\label{3.27}
\end{eqnarray}
The $B_{0}$ matrix is
\begin{eqnarray}
(B_{0})_{k,k} &=& -\lambda(\lambda +4k+1) + X_{k} :=Z_{k}(\lambda) ,\label{3.28}\\
(B_{0})_{k,k+1} &=& 2(k+1)^{2} .
\label{3.29}
\end{eqnarray}
The matrix $E$ is still not needed for $n=1$.

We define the matrix
\begin{equation}
(H_{1})_{k,r} = - (A_{1}^{-1}B_{0})_{k,r} =  [2(\lambda-1)]^{-1}\{ \beta_{k,r}(B_{0})_{r,r}
+ \beta_{k,r-1}(B_{0})_{r-1,r} \} ,\label{3.30}
\end{equation}
and obtain for the coefficients $ c_{k,1}^{(\alpha_{0})}$
\begin{equation}
c_{k,1}^{(\alpha_{0})} = \sum_{r=0}^{k+1} (H_{1})_{k,r} \left(-\frac{1}{2}\right)^{r} {\ell \choose r} .
\label{3.31}
\end{equation}
All these coefficients inherit a pole in $\lambda$ at the value $1$.

This pole does not appear in one eigenvalue only as in the $\zeta = \infty$ case. This is due to
the fact that for $\lambda =1$ there exist $ \ell+1$ degenerate regular solutions and therefore the pole appears in all $\ell+1$ eigenvalues simultaneously. It is straightforward to calculate the residues of all matrix elements of $H_{1}$ and to derive the expressions
\begin{equation}
\rho_{k} = \sum_{r=0}^{k+1} res (H_{1})_{k,r} \left( -\frac{1}{2}\right)^{r}{\ell\choose r} .
\label{3.32}
\end{equation}
Then we subtract from this solution at $n=1$ the regular solution
\begin{equation}
(\lambda -1)^{-1} [\sum_{r=0}^{\ell} \rho_{r} \Phi^{(r)}(u)] ,
\label{3.33}
\end{equation}
obtaining in the limit the log term of $\Psi^{(\ell)}[\Phi_{k}(u,\alpha_{0})]$
\begin{equation}
-\log u \quad[\sum_{r=0}^{\ell} \rho_{r} \Phi^{(r)}(u)] .
\label{3.34}
\end{equation}
We mention that the leading term of $\Psi^{(\ell)}[\Phi_{k}(u,\alpha_{0})]$ is
\begin{equation}
u^{-1}(I_{1} -\frac{1}{2} I_{2})^{\ell} .
\label{3.35}
\end{equation}

The situation with the Green function type solution is the same as with the solution which
is constrained by the AdS boundary condition: The UV constraint is satisfied by a coset, namely any linear combination of regular solutions can be added to the solution $\Psi^{(\ell)}[\Phi(\alpha_{0})]$. In turn this may also be used to normalize the
solutions $\Phi_{k}(\alpha_{0})$. We can namely require that on the level $n=1$ on which $\log{u}$ appears first, all the coefficients $c_{k,1}^{(\alpha_{0})}$ are made to vanish by appropriate subtraction of regular solutions.
\setcounter{equation}{0}
\section{Propagators in Feynman's gauge}
In this section we consider the higher spin gauge propagators analyzed in the previous section and in \cite{MR1}, \cite{LMR1}, \cite{LMR2} in an approach developed originally for the spin $\ell=0,1,2$ in \cite{Scal}, \cite{AllenJ}, \cite{Freed}
only, but now generalized for all $\ell$ with a slight modification of arguments. Namely we consider our propagator working directly in the space of conserved currents
\begin{equation}\label{4.1}
h^{(\ell )}(z_{1};a)= \int\sqrt{g}d^{4}z_{2}K^{(\ell)}(z_{1},a;z_{2},c)*_{c}J^{(\ell)}(z_{2},c) ,
\end{equation}
where
\begin{equation}\label{4.2}
  K^{(\ell)}(z_{1},a;z_{2},c)=\Psi^{(\ell)}[F_{k}(u(z_{1};z_{2}))] + \textnormal{traces} .
\end{equation}
Taking into account the conservation properties of the current $J^{(\ell)}(z_{2},c)$ we can formulate the ansatz following from (\ref{1.23})
\begin{eqnarray}
    [\Box_{1} +\ell-\Delta_{\ell}(\Delta_{\ell}-d)]\Psi^{(\ell)}[F_{k}(u(z_{1};z_{2}))]&=&-I^{\ell}_{1}\delta_{d+1}(z_{1};z_{2})+ \textnormal{traces}  \nonumber\\ &+&(c\nabla_{2})\left(I_{1a}\Psi^{(\ell-1)}[\Lambda_{k}(u(z_{1};z_{2}))]\right) .\qquad\quad\label{4.3}
\end{eqnarray}
This means that applying the gauge fixed equation of motion at the first argument of the bilocal propagator we get zero (or more precisely a delta function in the coincident points) due to a gauge transformation at the second argument.

Here we should make some comments on the delta function in curved $AdS$ space. Our notation in (\ref{4.3}) means
\begin{equation}\label{4.4}
   \delta_{(d+1)}(z_{1};z_{2})= \frac{\delta_{(d+1)}(z_{1}-z_{2})}{\sqrt{g(z)}} ,\quad\quad\quad\quad \int\delta_{(d+1)}(z_{1}-z_{2})d^{d+1}z_{1}=1 .
\end{equation}
In the polar coordinate system defined in Appendix A the invariant
measure (for $d=3$) is
\begin{equation}\label{4.5}
\sqrt{g}d^{4}z=u(u+2)dud\Omega_{3} .
\end{equation}
Therefore we can define
\begin{eqnarray}
  && \frac{\delta_{(4)}(z-z_{pole})}{\sqrt{g(z)}}=\frac{\delta(u)}{u(u+2)\Omega_{3}}
  =-\frac{\delta^{(1)}(u)}{(u+2)\Omega_{3}} ,\label{4.6}\\
  && u\delta^{(1)}(u)=-\delta(u)\nonumber .
\end{eqnarray}
This $u$- dependence of the measure leads to the idea that short distance singularities in $D=d+1=4$ dimensional $AdS$ space should start from
$\frac{1}{u^{2}}$ not from $\frac{1}{u}$.

Then using the gradient map (\ref{grad1}), (\ref{grad2}) we can derive
\begin{eqnarray}\label{4.7}
(c\nabla_{2})\left(I_{1a}\Psi^{(\ell-1)}[\Lambda_{k}(u)]\right)=\Psi^{(\ell)}[\Lambda'_{k-1}(u)+(k+1)\Lambda_{k}(u)], \quad\Lambda_{\ell}=0
\end{eqnarray}
Combining this with the Laplacian map (\ref{lm1})-(\ref{lm3}) and (\ref{4.1}) we obtain the following set of $\ell+1$ equations for $z_{1}\neq z_{2}$ (unlike the case (\ref{2.01}) we do not insert the value of $\Delta_{\ell}$ here)
\begin{eqnarray}
  && u(u +2)F_{k}'' +(d+1+4k)(u+1)F_{k}'+2(\ell-k+1)F_{k-1}'+2(u+1)(k+1)^{2}F_{k+1}\nonumber
\\&&+[2\ell+k(d+2\ell-k)]F_{k}-\Delta_{\ell}(\Delta_{\ell}-d)F_{k} = \Lambda'_{k-1}+(k+1)\Lambda_{k} .\label{4.8}
\end{eqnarray}
To analyze this system we write the $k=0,1$ and $\ell-1, \ell$ cases explicitly
\begin{eqnarray}
 && u(u +2)F_{0}''+(d+1)(u+1)F_{0}'-\Delta_{\ell}(\Delta_{\ell}-d)F_{0}+2(u+1)F_{1}+2\ell F_{0} = \Lambda_{0} ,\label{4.9}\\
 && \hspace{5.5cm} O(F_{1}'',F_{1}', F_{1}, F_{2})+2\ell F_{0}'= \Lambda'_{0}+2\Lambda_{1} ,\label{4.10}\\
&&  \hspace{5.5cm} \vdots  \nonumber\\
&&\hspace{5.5cm} O(F_{\ell-1}'',F_{\ell-1}', F_{\ell-1}, F_{\ell}, F_{\ell-2}') = \Lambda'_{\ell-2}+\ell\Lambda_{\ell-1} ,\quad\label{4.11}\\
&& u(u +2)F_{\ell}''+(d+1+4\ell)(u+1)F_{\ell}'+[\ell^{2}+\ell(d+2) -\Delta_{\ell}(\Delta_{\ell}-d)]F_{\ell}\nonumber\\
&&\hspace{10.5cm}+2F'_{\ell-1} = \Lambda'_{\ell-1} ,\label{4.12}
\end{eqnarray}
and we see that this system for $2\ell+1$ functions is separable. One solution is obtained if we put
\begin{eqnarray}
  F_{k} &=& 0 , \quad k=1,2,\dots \ell ,\label{4.13}\\
  \Lambda_{k}&=& 0 , \quad k=1,2,\dots \ell-1 ,\label{4.14}
\end{eqnarray}
and submit $F_{0}(u)$ to the Gaussian hypergeometric equation
 \begin{equation}\label{4.15}
   u(u +2)F_{0}''(u) +(d+1)(u+1)F_{0}'(u)-\Delta_{\ell}(\Delta_{\ell}-d)F_{0}(u)=0 ,
 \end{equation}
supplemented with a noncontradictory solution for the remaining gauge parameter $\Lambda_{0}(u)$
\begin{equation}\label{4.16}
    \Lambda_{0}(u)=2\ell F_{0}(u) .
\end{equation}

So we prove that with an appropriate choice of the gauge freedom we can obtain the propagator in Feynman's gauge in the form
\begin{equation}\label{4.17}
    K^{(\ell)}(z_{1},a;z_{2},c)= I^{\ell}_{1}F_{0}(u)+ \textnormal{traces} ,
\end{equation}
where $F_{0}(u)$ is the solution of the equation for the scalar field with dimension $\Delta_{\ell}$ (\ref{4.15}) \cite{Scal}. The solution of this equation is well known and can be written in two different forms \cite{Freed, FFF}. The first form is ($\zeta=u+1$)
\begin{equation}\label{4.18}
    F_{0}(\zeta)=C(\ell,d) 2^{\Delta_{\ell}} \zeta^{-\Delta_{\ell}}{}_{2}F_{1}\left(\frac{\Delta_{\ell}}{2},
    \frac{\Delta_{\ell}+1}{2},\Delta_{\ell}-\frac{d}{2}+1;\frac{1}{\zeta^{2}}\right) .
\end{equation}
This form is suitable for an investigation of the infrared behaviour. We see immediately that near the boundary limit we have
\begin{equation}\label{4.19}
     F_{0}(\zeta) \sim \zeta^{-\Delta_{\ell}}|_{d=3}=\zeta^{-(\ell+1)},   \quad\textnormal{if} \quad\zeta \rightarrow \infty ,
\end{equation}
which is just wanted for AdS/CFT correspondence. Indeed comparing  $\Delta_{\ell}$ and $\Delta_{\theta}$ in (\ref{1.23})-(\ref{1.25})  we see that the propagator of the nonphysical mode $\theta$ falls off in the boundary limit faster than the propagator for the physical mode $\psi$, as it should be.

But for us  the second form of this expression obtained after a quadratic transformation of the hypergeometric function listed in the Appendix B (\ref{B1}) is more interesting
\begin{equation}\label{4.20}
     F_{0}(u)=C(\ell,d) \left(\frac{2}{u}\right)^{\Delta_{\ell}}{}_{2}F_{1}\left(\Delta_{\ell},
   \Delta_{\ell} -\frac{d}{2}+\frac{1}{2},2\Delta_{\ell}-d+1;-\frac{2}{u}\right) .
\end{equation}
The normalization constant $C(\ell,d)$ is chosen to obtain the $\delta$ function on the right hand side of (\ref{4.3})
\begin{equation}\label{4.21}
    C(\ell,d)=\frac{\Gamma(\Delta_{\ell})\Gamma(\Delta_{\ell}
    -\frac{d}{2}+\frac{1}{2})}{(4\pi)^{\frac{(d+1)}{2}}\Gamma(2\Delta_{\ell}-d+1)}|_{d=3}=\frac{\ell!(\ell-1)!}{16\pi^{2}(2\ell-1)!} .
\end{equation}

To investigate the ultraviolet limit of (\ref{4.20}) we have to use the second formula  (\ref{B2}) of Appendix B and take carefully the limit $d\rightarrow 3$ to obtain
\begin{eqnarray}
  &&\left(\frac{2}{u}\right)^{\Delta_{\ell}}{}_{2}F_{1}\left(\Delta_{\ell},
   \Delta_{\ell} -\frac{d}{2}+\frac{1}{2},2\Delta_{\ell}-d+1;-\frac{2}{u}\right)|_{d\rightarrow 3}=\frac{(2\ell-1)!}{(\ell-1)!}\left\{\frac{2}{\ell! u}\right.\quad\quad\quad\quad\nonumber\\
  &&\left.+\frac{1}{(\ell-2)!}\sum^{\ell-2}_{n=0}\frac{(\ell+1)_{n}(2-\ell)_{n}}{n!(n+1)!}
  \left[\Upsilon_{\ell,n}+\log{\frac{u}{2}}\right]\left(-\frac{u}{2}\right)^{n}\right\} ,\label{4.22}
\end{eqnarray}
where the rational number $\Upsilon_{\ell,n}$ is expressed by the $\psi$ functions
\begin{equation}
   \Upsilon_{\ell,n}=\psi(\ell+n+1)+\psi(\ell-n-1)-\psi(n+1)-\psi(n+2) .
\end{equation}

So we see now that in the ultraviolet limit we get
\begin{equation}\label{4.23}
    F_{0}(u)|_{d=3}\cong\frac{1}{8\pi^{2}}\frac{1}{u} + O(1,u,\log{u},u\log{u}, \dots) .
\end{equation}
This main singular term in the propagator of the scalar field with dimension $\Delta_{\ell}$ does not depend on the field dimension and behaves always like $\frac{1}{8\pi^{2}u}$. For example we have the same singularity in the propagator of the conformally coupled scalar in $AdS_{4}$
(see \cite{MR3})
\begin{eqnarray}
  \Sigma[u(z_{1},z_{2})] &=&\frac{1}{8\pi^{2}}\left(\frac{1}{u}\pm \frac{1}{u+2}\right)  ,\label{4.24}\\
  (\Box+2)\Sigma[u(z_{1},z_{2})]&=& -\delta_{(4)}(z_{1};z_{2}) .\label{4.25}
\end{eqnarray}
So we observe some universality in  the UV behaviour of higher spin propagators in Feynman's gauge:

\emph{For any spin $\ell$ the propagator starts from $I^{\ell}_{1}\frac{1}{8\pi^{2}u}$.}

Comparing with (\ref{3.35}) we deduce that in de Donder gauge we have the same picture because
\begin{itemize}
  \item $I_{1}(a,c;u)\rightarrow a^{\mu}c_{\nu}$
  if $u\rightarrow 0$ .
  \item $I_{2}(a,c;u)=I_{3}(a,c;u) \rightarrow 0$ if $u\rightarrow 0$ .
  \item $I_{4}(a,c;u)\rightarrow a^{2}c^{2}$ if $u\rightarrow 0$ .
\end{itemize}
So finally we can formulate the following statement:

\emph{The higher spin propagator in Feynman's gauge  is simplest and most convenient for the calculation of any Feynman diagram. Just we have to couple it with conserved currents to make sure that we preserve gauge invariance. The UV-behaviour of the propagator is universal and described by (\ref{4.23})}.

\setcounter{equation}{0}
\section{Spin $\ell$, $\ell-2$ and scalar interaction and mass renormalization}
Now we turn to consider some special cases of higher spin interaction. The first solutions for this old problem for $AdS$ space  was considered many years ago \cite{Vasiliev}. We will not pretend in these notes to refer all articles that appeared during last 30 years dealing with this important problem in general but eventually using different approaches. We select here only recent reviews in this field \cite{review} and note that some similar interactions were considered in \cite{Petkou} in a type of BRST approach \cite{brst}.

In this section we will construct all possible gauge invariant interactions between two neighboring higher spin  gauge fields and a scalar containing two derivatives. On this linearized level of understanding the higher spin gauge invariance it is possible to construct an interaction of the gauge field contracted with the conserved current formed from gauge fields of the nearest different spin ($\ell\pm 2$), comformally coupled in the $AdS_{d+1}$ background with the scalar $\sigma(z)$
\begin{equation}\label{5.1}
   \Box \sigma(z)+\frac{d^{2}-1}{4}\sigma(z)=0 ,
\end{equation}
and two derivatives
\begin{equation}\label{5.2}
    S_{int}=\frac{g_{\ell}}{\sqrt{N}}\int \sqrt{g}d^{4}z h^{(\ell)}(z;a)*_{a}J^{(\ell)}[h^{(\ell\pm 2)}(z;a),\sigma(z)] .
\end{equation}
Here we introduce an unknown coupling parameter $g_{\ell}$ normalized as $O(\frac{1}{\sqrt{N}})$ as it follows from $AdS_4$/$CFT_{3}$ correspondence.
The conservation condition following from the gauge transformation for $h^{(\ell)}(z)$ (\ref{1.7}) with the \emph{traceless} parameter $\epsilon^{(\ell-1)}$ is
\begin{equation}\label{5.3}
    \nabla^{\mu}\frac{\partial}{\partial a_{\mu}}J^{(\ell)}[h^{(\ell\pm 2)}(z;a),\sigma(z)]= O(a^{2}) .
\end{equation}
This equation  could be used to construct all possible currents with  properties mentioned above. Actually in our previous article \cite{MR2} we constructed a conserved current solving the conservation condition directly for the case of spin $\ell-2$, a scalar field and two gradient derivatives. Moreover we have shown there that the solution is unique and should start from the double full derivative term. Here we will present a more general way of solving the conservation condition (\ref{5.3}). Note also that in this way we can construct only an interaction of scalars with two different higher spin fields transforming with different gauge parameters. For the construction of the interaction of the scalar with the selfinteracting
gauge field $h^{(\ell)}h^{(\ell)}\sigma$ we need more than the linearized theory of higher spins. In this case we should analyze and construct first the term quadratic in $h^{(\ell)}$  of the gauge invariant Fronsdal operator deforming at the same time the gauge transformation up to a form linear in $h^{(\ell)}$ . This could be done in principle in an $AdS$ background at least and we postpone this task for future investigations.

Returning to the equation (\ref{5.3}) we note first that the operator
\begin{equation}\label{5.4}
\mathcal{\widetilde{D}}^{(+1)}=(a\nabla)-\frac{1}{2}a^{2}\nabla ^{\mu }\frac{\partial }{\partial a^{\mu }}
\end{equation}
is dual  to the de Donder gauge operator
\begin{eqnarray}
&&\mathcal{D}^{(-1)}=\nabla ^{\mu }\frac{\partial }{\partial a^{\mu }}-\frac{1}{2}(a\nabla )\Box _{a},
\label{5.5}
\end{eqnarray}
with respect to the full scalar product (\ref{1.15}),
and second that this operator commutes with the divergence in the following way (see (\ref{A.7}) and (\ref{A.8}))
 \begin{equation}\label{5.6}
    \nabla ^{\mu }\frac{\partial }{\partial a^{\mu }}\mathcal{\widetilde{D}}^{(+1)}j^{(\ell-1)}(z,a)=[\Box-(\ell-1)(d+\ell-2)]j^{(\ell-1)}(z,a) +O(a^{2}) .
 \end{equation}
Then taking into account (\ref{A.9}) we can see that with the natural choice $j^{(\ell-1)}(z,a)=(a\nabla)[h^{(\ell-2)}(z;a)\sigma(z)]$
one can obtain ($\mu^{2}_{\theta^{(\ell-2)}}$ is the $AdS$ mass of the trace part of $h^{(\ell)}$ (\ref{1.21}))
\begin{equation}\label{5.7}
    \nabla ^{\mu }\frac{\partial }{\partial a^{\mu }}\mathcal{\widetilde{D}}^{(+1)}(a\nabla)[h^{(\ell-2)}(z;a)\sigma(z)]
    =(a\nabla)[\Box-\mu^{2}_{\theta^{(\ell-2)}}](h^{(\ell-2)}(z;a)\sigma(z))+O(a^{2}) .
\end{equation}
This can be integrated easily and we restore our conserved current from \cite{MR2} in the following form
\begin{equation}\label{5.8}
    J^{(\ell)}[h^{(\ell-2)},\sigma]=\mathcal{\widetilde{D}}^{(+1)}(a\nabla)[h^{(\ell-2)}(z;a)\sigma(z)]
    -\frac{a^{2}}{2}[\Box-\mu^{2}_{\theta^{(\ell-2)}}](h^{(\ell-2)}(z;a)\sigma(z))+O(a^{4}) .
\end{equation}
Note that all $O(a^{4})$ terms are unimportant due to the double tracelessness of $h^{(\ell)}(z;a)$.
At this point we will apply for simplicity de Donder's gauge condition to all types of gauge fields.
Then using free equations of motion only for the fields $h^{(\ell-2)}$ and $\sigma$ that form the conserved current, and neglecting the first part due to de Donder's gauge condition for the gauge field $h^{(\ell)}$, we obtain the following effective current
\begin{equation}\label{5.9}
    J^{(\ell)}[h^{(\ell-2)}, \sigma]=-\frac{a^{2}}{2}\left[2\nabla^{\mu}(\nabla_{\mu}h^{(\ell-2)}\sigma)
    +\left(\frac{1-d^{2}}{4}-\mu^{2}_{(\ell-2)}-\mu^{2}_{\theta^{(\ell-2)}}\right)h^{(\ell-2)}\sigma\right] .
\end{equation}

Note that this interaction vanishes if we require a free equation of motion for  the field $h^{(\ell)}$ coupled to the conserved current.

The next step of our consideration is the construction of the conserved current $ J^{(\ell)}[h^{(\ell+2)},\sigma]$ which is dual to the former one, where the gauge field inside the current has a spin higher than the gauge field coupled with the current.
Exploring in a similar way the conservation condition (\ref{5.3}) and using divergence instead of gradient on stage (\ref{4.7}) and formula (\ref{A.6}) instead of (\ref{A.9}) we obtain the following solution
\begin{equation}\label{5.10}
 J^{(\ell)}[h^{(\ell+2)},\sigma]=\mathcal{\widetilde{D}}^{(+1)}(\nabla \partial_a )[\theta^{(\ell)}(z;a)\sigma(z)]
    -[\Box-\mu^{2}_{(\ell)}](\theta^{(\ell)}(z;a)\sigma(z)) ,
\end{equation}
where
\begin{equation}\label{5.11}
    \theta^{(\ell)}(z;a)=\Box_{a}h^{(\ell+2)}(z;a) .
\end{equation}
Inserting in (\ref{5.10}) $\ell-2$ instead of $\ell$ and using the equation of motion for the fields inside the current and de Donder' gauge condition for the external gauge field, we obtain the effective current
\begin{equation}\label{5.12}
     J^{(\ell-2)}[h^{(\ell)},\sigma]= -\left[2\nabla^{\mu}(\nabla_{\mu}\theta^{(\ell-2)}\sigma)
    +\left(\frac{1-d^{2}}{4}-\mu^{2}_{(\ell-2)}-\mu^{2}_{\theta^{(\ell-2)}}\right)\theta^{(\ell-2)}\sigma\right] .
\end{equation}
Comparing with (\ref{5.9}) we see that in both cases we have the same effective interaction between the physical mode $\psi^{(\ell-2)}$, the trace mode $\theta^{(\ell-2)}$ and the scalar, and the conserved current (\ref{5.12}) up to an overall normalization is dual to the conserved current (5.9) due to the equations of motion for the fields forming the currents in each cases.
Thus we prove that one can use Feynman's gauge for propagators coupled to these two currents and can turn now to investigate some loop diagram for a study of mass renormalization or quantum mass generation phenomena. Actually we considered all interactions of two neighbouring higher spin fields with a Higgs scalar that are minimal with respect to the number of derivatives, and which can generate mass in a loop. Though from (\ref{5.6}) follows that we can introduce in principle many other $j^{(\ell-1)}(z;a)$, all of them will contain more derivatives  in front of the quantized fields and will generate counterterms that are not suitable for finite mass renormalization.

So we see that only one reasonable one loop  diagram can be constructed from the interactions considered in this section. It is a loop formed by the scalar $\sigma$ and the nonphysical trace mode $\theta^{(\ell)}$ and with physical but off-shell external lines $\psi^{(\ell)}$. Actually we would like to calculate the following quadratic part of the effective action
\begin{equation}\label{5.13}
    \frac{g^{2}_{\ell}}{N}\int \sqrt{g}d^{4}z_{1}\int \sqrt{g}d^{4}z_{2} h^{(\ell)}(z_{1};a)*_{a}\big< J^{(\ell)}[h^{(\ell+2)},\sigma;z_{1};a],J^{(\ell)}[h^{(\ell+2)},\sigma;z_{2};c)]\big>*_{c}h^{(\ell)}(z_{2};c) ,
\end{equation}
where $J^{(\ell)}[h^{(\ell+2)},\sigma;z_{2};c)]$ is presented in (\ref{5.10}). Performing a partial integration and taking into account tracelessness of $\theta^{(\ell)}$ we get the following expression
\begin{equation}\label{5.14}
     \frac{g^{2}_{\ell}}{N}\int \sqrt{g}d^{4}z_{1}\int \sqrt{g}d^{4}z_{2}\mathcal{F}(\psi^{(\ell )}(z_{1};a))*_{a}\Sigma[u(z_{1},z_{2})]\Theta^{(\ell)}[u(z_{1},z_{2});a,c]*_{c}\mathcal{F}(\psi^{(\ell )}(z_{2};c)) .
\end{equation}
Here $\mathcal{F}(\psi^{(\ell )})$ is the traceless part of Fronsdal's operator, $\Sigma[u]$ is the scalar propagator (\ref{4.24}) and
\begin{equation}\label{5.15}
    \Theta^{(\ell)}[u(z_{1},z_{2});a,c]=\big<\theta^{(\ell)}(z_1;a),\theta^{(\ell)}(z_2;c)\big>
\end{equation}
is a trace part of the $h^{(\ell+2)}$ propagator. We want to understand the singular part of this loop.

From now on we follow a technique developed in previous articles \cite{MR3} and \cite{MR4} where the trace anomaly of the scalar mode in external higher spin field was successfully calculated from a one loop diagram.
First we can use an $AdS$ transformation to fix the point $z_{1}$ as a
pole for the coordinate system $z_{2}$. Then the integration measure can be expressed through the chordal distance $u$ as it is explained in the Appendix A. The singularity of the product of the scalar and the higher spin propagators is relevant if it is at least $1/u^{2}$ because one power of $u$ is compensated by the integration measure (see (\ref{invv}) and explanation hereafter). Then from the relative coefficient between the $\psi$ and $\theta$ modes in (\ref{1.20}) evaluated for spin $\ell+2$ and $d=3$, from (\ref{4.23}) and an additional sign from the indefinite metric (\ref{1.26}) we deduce
 \begin{equation}\label{5.16}
   \big< \theta^{(\ell)}(z_1;a),\theta^{(\ell)}(z_2;c)\big>=-\frac{4(\ell+2)}{\ell+1}I^{\ell}_{1}\frac{1}{8\pi^{2}u} + O(u, \log{u},\dots) ,
 \end{equation}
Multiplying hereafter with the scalar propagator we get the unique singular term of the loop function
 \begin{equation}\label{5.17}
   \left\{\Sigma[u]\Theta^{(\ell)}[u]\right\}_{sing}=-\frac{(\ell+2)}{16\pi^{4}(\ell+1)}I^{\ell}_{1}\frac{1}{u^{2}} .
 \end{equation}
 Using a standard formula of analytic dimensional regularization in $AdS$ space (see \cite{MR3, MR4})
 \begin{equation}\label{5.18}
    \left[\frac{1}{u^{n-\epsilon}}\right]_{sing}=-
    \frac{1}{\epsilon}\frac{(-1)^{n-1}}{(n-1)!}\delta^{(n-1)}(u) .
 \end{equation}
for our distribution with $n=2$ and the definition of the delta function (\ref{4.6}), we obtain
\begin{equation}\label{5.19}
  \left\{\Sigma[u]\Theta^{(\ell)}[u]\right\}_{sing}=-\frac{\Omega_{3}(\ell+2)}{8\pi^{4}(\ell+1)}(a^{\mu}c_{\mu})^{\ell}\frac{1}{\epsilon} \delta_{(4)}(z_{1};z_{2}) .
\end{equation}

Before inserting this expression in (\ref{5.14}) we have to be sure that we preserved gauge invariance during regularization.
At this stage it means that we have to preserve the conservation condition (\ref{5.3}) for the current as a Ward identity for the correlator in (\ref{5.13}). Then taking into account that after partial integration we got gauge invariant Fronsdal's operators instead of external lines we deduce that we just should write these external lines as a gauge invariant object during dimensional regularization or in other words for  $d=3-\epsilon$. From the formula for the geometric $AdS$ mass $\mu^{2}_{\ell}$  (\ref{1.5}) we see that
\begin{equation}\label{5.20}
    \mathcal{F}^{d=3-\epsilon}(h^{(\ell)}(z;a))=\mathcal{F}^{d=3}(h^{(\ell)}(z;a))+\epsilon(\ell-2)h^{(\ell)}(z;a) .
\end{equation}
Then inserting this  and (\ref{5.19}) in (\ref{5.14}) we obtain immediately as local singularity of our diagram
\begin{equation}\label{5.21}
    -\frac{1}{\epsilon}\frac{g^{2}_{\ell}\Omega_{3}(\ell+2)}{8N\pi^{4}(\ell+1)}\int \sqrt{g}d^{4}z\mathcal{F}(\psi^{(\ell )}(z;a))*_{a}\mathcal{F}(\psi^{(\ell )}(z;a)) ,
\end{equation}
supplemented with the additional finite local term
\begin{equation}\label{5.22}
    -\frac{g^{2}_{\ell}\Omega_{3}(\ell-2)(\ell+2)}{4N\pi^{4}(\ell+1)}\int \sqrt{g}d^{4}z\psi^{(\ell )}(z;a)*_{a}\mathcal{F}(\psi^{(\ell )}(z;a)) .
\end{equation}

The first singular term can be dropped  adding the same  singular local and gauge invariant counterterm to the effective action as (\ref{5.21}) but with opposite sign. The second finite local part is not gauge invariant itself and  cannot be absorbed by adding the local finite invariant counterterm but can be absorbed by finite renormalization of the mass term. Indeed let us add an additional finite local counterterm proportional to
\begin{equation}\label{5.23}
    \int \sqrt{g}d^{4}z[\mathcal{F}(\psi^{(\ell )})-\delta m^{2}_{\ell} \psi^{(\ell )} ]*_{a}[\mathcal{F}(\psi^{(\ell )})-\delta m^{2}_{\ell} \psi^{(\ell )}] .
\end{equation}
Then we see that if
\begin{equation}\label{5.24}
    \delta m^{2}_{\ell} =\frac{g^{2}_{\ell}\Omega_{3}(\ell-2)(\ell+2)}{8N\pi^{4}(\ell+1)}=\frac{g^{2}_{\ell}(\ell-2)(\ell+2)}{4N\pi^{2}(\ell+1)} ,
\end{equation}
we will cancel (\ref{5.22}) without any additional term in the given order of perturbation theory ($O(1/N)$), absorbing the additional $O(1)$ finite local $\mathcal{F}^{2}$ term in the infinite singular \emph{gauge invariant} counterterm, since an additional finite renormalization supplementing the infinite one fixes the renormalization scheme. We see that our mass renormalization implies a soft symmetry breaking because we got only a finite mass generation. In other words all our infinite counterterms are gauge invariant.

So we see that we got mass renormalization as it was expected from $AdS_{4}/CFT_{3}$ correspondence and formulated in terms of boundary $CFT$ theory in an article of one of the authors \cite{Ruehl}. In principle we can compare this mass with the answer obtained in \cite{Ruehl} from anomalous dimensions of higher spin currents in the $O(N)$ sigma model
(\ref{0.1}).
Especially we got the same interesting $(\ell-2)$ factor protecting the spin 2 graviton field, that corresponds to the boundary energy momentum tensor, from renormalization and find a prediction for the coupling $g_{\ell}$. But we will not do that at this stage because it is not the full solution of the problem. As it was mentioned above we did not include in our consideration the selfinteraction $h^{(\ell)}h^{(\ell)}\sigma$ and the corresponding one loop diagram. To consider this we need a nonlinear (at least second order) formulation for higher spin curvatures and Fronsdal's operators which we hope we will be able to do in  the near future, continuing and generalizing our construction at linear order performed in \cite{MR5, MR6}. Then we will be able to compare this  UV approach with another IR ansatz including the St\"uckelberg and Goldstone mechanism which was considered in \cite{MR2}. Note that in \cite{MR2} we used another loop diagram with the $J^{(\ell)}(h^{(\ell-2)}\sigma)$ current and the protection of the graviton mode appeared there just due to the absence of that diagram for $\ell=2$.

\subsection*{Acknowledgements}
\quad This work was supported by Alexander von Humboldt Foundation under 3.4-Fokoop-ARM/1059429.
Work of K.M. also partially supported by CRDF-NFSAT UCEP06/07.

\section*{Appendix A}
\setcounter{equation}{0}
\renewcommand{\theequation}{A.\arabic{equation}}
The Euclidian $AdS_{d+1}$ metric
\begin{equation}\label{eads}
    ds^{2}=g_{\mu \nu }(z)dz^{\mu }dz^{\nu
}=\frac{1}{(z^{0})^{2}}\delta _{\mu \nu }dz^{\mu }dz^{\nu }
\end{equation}
can be realized as an induced metric for the hypersphere defined by
the following embedding procedure in $d+2$ dimensional Minkowski
space
\begin{eqnarray}
  && X^{A}X^{B}\eta_{AB}=-X_{-1}^{2}+X_{0}^{2}+\sum^{d}_{i=1}X_{i}^{2}=-1 ,\\
  && X_{-1}(z)=\frac{1}{2}\left(\frac{1}{z_{0}}+\frac{z_{0}^{2}
  +\sum^{d}_{i=1}z^{2}_{i}}{z_{0}}\right) ,\\
  && X_{0}(z)=\frac{1}{2}\left(\frac{1}{z_{0}}-\frac{z_{0}^{2}
  +\sum^{d}_{i=1}z^{2}_{i}}{z_{0}}\right) ,\\
  && X_{i}(z)=\frac{z_{i}}{z_{0}} .
\end{eqnarray}
Using these embedding rules we can identify the variable
$\zeta(z,w)$ as an $SO(1,d+1)$ invariant scalar product
\begin{equation}\label{gd}
    -X^{A}(z)Y^{B}(w)\eta_{AB}=\frac{1}{2z_{0}w_{0}}\left(2z_{0}w_{0}
    +\sum^{d}_{\mu=0}(z-w)^{2}_{\mu}\right)=\zeta=u+1 ,
\end{equation}
and therefore can be realized by $cosh$ of a hyperbolic angle. Indeed we can
introduce another embedding
\begin{eqnarray}\label{hyp}
    &&X_{-1}(\eta,\omega_{\mu})=\cosh{\eta},\\
   && X_{\mu}(\eta,\omega_{\mu})=\sinh{\eta}\,\omega_{\mu}\quad,\quad\quad
    \sum^{d}_{\mu=0}\omega^{2}_{\mu}=1 ,\\
    &&ds^{2}=d\eta^{2}+\sinh^{2}{\eta}\, d\Omega_{d} .
\end{eqnarray}
In these coordinates the chordal distance $u$ between an arbitrary point
$X^{A}(\eta,\Omega_{\mu})$ and the pole of the hypersphere
$Y^{A}(\eta=0,\omega_{\mu})$ is simply
\begin{equation}\label{hd}
    \zeta= -X^{A}Y^{B}\eta_{AB}=\cosh{\eta} .
\end{equation}
Therefore the invariant measure is expressed as
\begin{equation}\label{invv}
    \sqrt{g}d\eta d\Omega_{d}=(\sinh\eta)^{d}d\eta
    d\Omega_{d}=[u(u+2)]^{\frac{d-1}{2}}du d\Omega_{d} .
\end{equation}
So we see that the integration measure for $d=3$ ($D=d+1=4$) will
cancel one order of $u^{-n}$ in short distance singularities and we
have to count the singularities starting from $u^{-2}$ which is  "logarithmically divergent"
in standard QFT terminology.

 In this article we use the following rules and relations for
$u(z,z')$, $I_{1a}$, $I_{2c}$ and the bitensorial basis
$\{I_{i}\}^{4}_{i=1}$
\begin{eqnarray}
  && \Box u=(d+1)(u+1) ,\quad \nabla_{\mu}\partial_{\nu}u=g_{\mu\nu}(u+1) ,
  \quad g^{\mu\nu}\partial_{\mu}u\partial_{\nu}u=u(u+2) ,\quad\quad\quad\label{start}\\
  &&   \partial_{\mu}\partial_{\nu'}u
  \nabla^{\mu}u=(u+1)\partial_{\nu'}u ,\quad
  \partial_{\mu}\partial_{\nu'}u \nabla^{\mu}\partial_{\mu'}u
  =g_{\mu'\nu'}+\partial_{\mu'}u\partial_{\nu'}u ,\\
&&\nabla_{\mu}\partial_{\nu}\partial_{\nu'}u \nabla^{\mu}u
  =\partial_{\nu}u\partial_{\nu'}u ,\quad
  \nabla_{\mu}\partial_{\nu}\partial_{\nu'}u
  =g_{\mu\nu}\partial_{\nu'}u ,\\
&&\frac{\partial}{\partial a^{\mu}}I_{1a}\frac{\partial}{\partial
a_{\mu}}I_{1a}=u(u+2) ,\quad \frac{\partial}{\partial
a^{\mu}}I_{1}\frac{\partial}{\partial
a_{\mu}}I_{1a}=(u+1) I_{2c} ,\\
&&\frac{\partial}{\partial a^{\mu}}I_{1}\frac{\partial}{\partial
a_{\mu}}I_{1}=c^{2}_{2}+ I_{2c}^{2} , \, \frac{\partial}{\partial
a^{\mu}}I_{1}\frac{\partial}{\partial a_{\mu}}I_{2}=(u+1) I_{2c}^{2}
,\,\Box_{a}I_{4}=2(d+1)c^{2}_{2} ,\\
 &&\frac{\partial}{\partial
a^{\mu}}I_{2}\frac{\partial}{\partial a_{\mu}}I_{2}=u(u+2)I_{2c}^{2}
,\quad
\Box_{a}I_{3}=2(d+1)I_{2c}^{2}+2c^{2}_{2}u(u+2) ,\\
&&\nabla^{\mu}\frac{\partial}{\partial a^{\mu}}I_{1}=(d+1)I_{2c}
,\,\nabla^{\mu}\frac{\partial}{\partial a^{\mu}}I_{2}=(d+2)(u+1)
I_{2c},\quad\nabla^{\mu} I_{1}\partial_{\mu}u=I_{2} ,\\
&&\nabla^{\mu}\frac{\partial}{\partial
a^{\mu}}I_{3}=4I_{1}I_{2c}+2(d+2)(u+1) c^{2}_{2}I_{1a}
,\quad\nabla^{\mu} I_{2}\partial_{\mu}u=2(u+1) I_{2} ,\\
&&\frac{\partial}{\partial a_{\mu}} I_{1}\partial_{\mu}u=(u+1)
I_{2c} ,\quad \frac{\partial}{\partial a_{\mu}}
I_{2}\partial_{\mu}u=u(u+2) I_{2c} ,\,\frac{\partial}{\partial
a_{\mu}}
I_{1}\nabla_{\mu} I_{1}=I_{1} I_{2c} ,\,\,\,\,\,\quad\quad\\
&&\frac{\partial}{\partial a_{\mu}} I_{1}\nabla_{\mu}
I_{2}=I_{2c}\left((u+1) I_{1}+I_{2}\right)+c^{2}_{2}I_{1a}
,\frac{\partial}{\partial a_{\mu}} I_{2}\nabla_{\mu}
I_{1}=I_{2c}I_{2} ,\\
&&\frac{\partial}{\partial a_{\mu}} I_{2}\nabla_{\mu} I_{2}=2(u+1)
I_{2c}I_{2} ,\quad \nabla^{\mu} I_{1}\nabla_{\mu}
I_{1}=a^{2}_{1}I_{2c}^{2} ,\quad \Box I_{1}=I_{1} ,\\
&&\nabla^{\mu} I_{1}\nabla_{\mu} I_{2}=I_{2}I_{1}+ a^{2}_{1}(u+1)
I_{2c}^{2} ,\quad \Box I_{2}=(d+2)I_{2}+2(u+1)
I_{1} ,\quad\\
&&\nabla^{\mu} I_{2}\nabla_{\mu} I_{2}=I_{2}^{2}+2(u+1)
I_{1}I_{2}+a^{2}_{1}I_{2c}^{2}(u+1)^{2}+c^{2}_{2}I_{1a}^{2}
,\\
&&a^{\mu}\nabla_{\mu}I_{1a}=a^{2}(u+1) ,\quad
a^{\mu}\nabla_{\mu}I_{2c}=I_{1},\quad
a^{\mu}\nabla_{\mu}I_{1}=a^{2}I_{2c},
\\&&a^{\mu}\nabla_{\mu}I_{2}=a^{2}(u+1) I_{2c}+I_{1a}I_{1},
.\label{end}
\end{eqnarray}
Using these relations we can derive ($
F'_{k}:=\frac{\partial}{\partial
  u}F_{k}(u)$)
\begin{itemize}
  \item Divergence map
\begin{eqnarray}
  && \nabla^{\mu}_{1}\frac{\partial}{\partial a^{\mu}}\Psi^{\ell}[F]=
  I_{2c}\Psi^{\ell-1}[Div_{\ell}F]+O(c^{2}_{2}) , \label{div}\\
  && (Div_{\ell}F)_{k}=(\ell-k)(u+1)
  F'_{k}+(k+1)u(u+2)F'_{k+1}\nonumber\\
  &&+(\ell-k)(\ell+d+k)F_{k}+(k+1)(\ell+d+k+1)(u+1) F_{k+1} .\label{dv}
  \end{eqnarray}
  \item Trace map
\begin{eqnarray}
    &&\Box_{a}\Psi^{\ell}[F]=I^{2}_{2c}\Psi^{\ell-2}
    [{Tr_{\ell}F}]+O(c^{2}_{2}) ,\label{tracemap}\\
&&(Tr_{\ell}F)_{k}=(\ell-k)(\ell-k-1)F_{k} +2(k+1)(\ell-k-1)(u+1)
F_{k+1}\nonumber\\&&\quad\quad\quad\quad\quad+(k+2)(k+1)u(u+2)F_{k+2} .\label{tr}
\end{eqnarray}
  \item Laplacian map
  \begin{eqnarray}
  && \Box_{1} \Psi^{\ell}[F]=\Psi^{\ell}[Lap_{\ell}F]+O(a^{2}_{1},c^{2}_{2}) , \label{lm1}\\
  &&(Lap_{\ell}F)_{k}=u(u+2)F''_{k}+(d+1+4k)(u+1)
  F'_{k}+[\ell+k(d+2\ell-k)]F_{k}\nonumber\\&&+2(u+1)(k+1)^{2}
  F_{k+1}+2(\ell-k+1)F'_{k-1},\label{lm2}\\
  &&\Box F_{k}(u)=u(u+2)F''_{k}+(d+1)(u+1) F'_{k}.\quad\label{lm3}\end{eqnarray}
  \item Gradient map
  \begin{eqnarray}
&&(a\cdot\nabla)_{1}\Psi^{\ell}[F]=I_{1a}\Psi^{\ell}[Grad_{\ell}F]+ O(a^{2}_{1}) ,\label{grad1}\\
  && (Grad_{\ell}F)_{k}=F'_{k}+(k+1)F_{k+1} .\label{grad2}
\end{eqnarray}

\end{itemize}

At the end we present all important commutation relations working in the space of symmetric rank $n$ tensors
\begin{eqnarray}
  &&[(\nabla\partial_{a}), \Box]f^{(n)}(z,a)= \left[2(a\nabla)\Box_{a}-(d+2n-2)(\nabla\partial_{a})\right]f^{(n)}(z,a) ;\quad \quad\label{A.6}\\
  &&[(\nabla\partial_{a}), (a\nabla)]f^{(n)}(z,a)= \Box f^{(n)}(z,a)
 + [\nabla_{\mu}, (a\nabla)]\partial_{a}^{\mu}f^{(n)}(z,a) ;\label{A.7}\\
&& [\nabla_{\mu}, (a\nabla)]\partial_{a}^{\mu}f^{(n)}(z,a)= \left[a^{2}\Box_{a}-n(d+n-1)\right]f^{(n)}(z,a) ;\label{A.8}\\
&&[\Box ,(a\nabla)]f^{(n)}(z,a)= \left[2a^{2}(\nabla\partial_{a})-(d+2n)(a\nabla)\right]f^{(n)}(z,a) ;\label{A.9}\\
&&\Box_{a}\left[a^{2}f^{(n)}(z,a)\right] = 2(d+2n+1)f^{(n)}(z,a)+a^{2}\Box_{a}f^{(n)}(z,a) .\label{A.10}
\end{eqnarray}
\section*{Appendix B}
\setcounter{equation}{0}
\renewcommand{\theequation}{B.\arabic{equation}}
These two useful hypergeometric identities we learned from the book of H. Bateman  and A. Erdelyi ``Higher transcendental functions'' V.1, McGraw-Hill Book company Inc. 1953.
\begin{eqnarray}
   _{2}F_{1}(a,b,2b;z)&=& \left(1-\frac{z}{2}\right)^{-a} {}_{2}F_{1}\left(\frac{a}{2},\frac{a+1}{2},b+\frac{1}{2};\left(\frac{z}{2-z}\right)^{2}\right) ,\label{B1}\\
  {}_{2}F_{1}(a,b,c;z)&=&\frac{\Gamma(c)\Gamma(b-a)}{\Gamma(b)\Gamma(c-a)}(-z)^{-a} {}_{2}F_{1}(a,1-c+a,1-b+a;z^{-1})\nonumber\\
  &+&\frac{\Gamma(c)\Gamma(b-a)}{\Gamma(b)\Gamma(c-a)}(-z)^{-b} {}_{2}F_{1}(a,1-c+b,1-a+b;z^{-1}) .\label{B2}
\end{eqnarray}

\end{document}